\documentclass[12pt]{article}
\usepackage{amssymb,amsmath,epsfig}
\usepackage{cite}
\allowdisplaybreaks

\begin{document}
\title{\bf Definition of Complexity Factor for Self-Gravitating Systems in Palatini $f(R)$ Gravity}

\author{Z. Yousaf
\thanks{zeeshan.math@pu.edu.pk}\\
Department of Mathematics, University of the Punjab,\\
Quaid-i-Azam Campus, Lahore-54590, Pakistan.}
\date{}
\maketitle

\begin{abstract}
The aim of this paper is to explore the complexity factor for those
self-gravitating relativistic spheres whose evolution proceeds non-
dynamically. We are adopting the definition of CF mentioned in
\cite{PhysRevD.97.044010}, modifying it to the static spherically
symmetric case, within the framework of a modified gravity theory
(the Palatini $f(R)$ theory). In this respect, we have considered
radial dependent anisotropic matter content coupled with spherical
geometry and determined the complexity factor involved in the
patterns of radial evolution. We shall explore the field and a
well-known Tolman-Oppenheimer-Volkoff equations. After introducing
structure scalars from the orthogonal decomposition of the Riemann
tensor, we shall calculate complexity factor. An exact analytical
model is presented by considering firstly ansatz provided by Gokhroo
and Mehra. The role of matter variables and $f(R)$ terms are
analyzed in the structure formation as well as their evolution
through a complexity factor.
\end{abstract}

\section{Introduction}

One of the most important gravitational theories, presented by
Albert Einstein about more than a century ago is general relativity
(GR). This theory has rejected Newton's idea of gravitational
force and described the deformation of space-time geometry through
the mass-energy distribution. The recent plethora of observations,
like, type Ia supernovae, large scale
structure and cosmic microwave background radiation \cite{pietrobon2006integrated,giannantonio2006high,riess2007new},
states that our cosmos is characterized by an accelerated
expansion. Furthermore, various recent cosmic observations at astronomical
scales \cite{ostriker1973numerical,refregier2003weak}
confirmed the unknown nature of the most matter part of the universe.

After the observational works presented by LIGO/Virgo collaboration
for understanding neutron star mergers \cite{abbott2017gw170817},
binary compact systems
\cite{abbott2016observation,abbott2017gw170814}, and the initiative
probes of EUCLID \cite{laureijs2011euclid}, the cosmic models
provided by modified theories of gravity (MTG) will be put to
experimental verification in extragalactic and cosmological
backgrounds. Many cosmological models of MGT have been ruled out by
current observational outcomes from the mergers of neutron stars.
Not only this, the severe restriction has also been imposed by such
experiments on the viability of some MTG
\cite{lombriser2017challenges,baker2017strong}. There has been a
number of achievements of GR, however, the search for detection of
dark matter/energy sources in view of the concordance model,
space-time singularities \cite{senovilla20151965}, etc may require
some MTG. Keeping in mind the presence of unobserved (up to this
time) dark matter (DM) and dark energy (DE), a new route of research
was suggested based on the generalizations of GR. This famous idea
has received standard terminology called MTG.

Qadir et al. \cite{qadir2017modified} also suggested that GR may
need to extend in order to study the various aspects of gravitation
with quantum cosmology. There has been very interesting cosmological
models related to MTG for the description of stellar collapse and
their evolution
\cite{capozziello2010beyond,capozziello2011extended,bamba2012dark,doi:10.1142/S021773231950189X,sahu2017cosmic}.
Nojiri and Odintsov \cite{nojiri2007introduction} described not only
the significance of MGT but also elaborated the applications of such
theories in the discussion of comic evolution. The straightforward
MTG from GR is $f(R)$ ($R$ is the Ricci scalar)
\cite{capozziello2011extended,elizalde2011nonsingular} and
$f(\mathcal{T})$ ($\mathcal{T}$ is the torsion scalar)
\cite{bamba2010cosmological}. After which, such theories are
generalized by including amalgams of curvature terms in a more
complex ways, for instance, $f(R,\Box R, T)$ ($\Box$ is the de
Alembert's operator and $T$ is the trace of energy momentum tensor)
\cite{houndjo2017higher,doi:10.1142/S0219887818501463}, $f(G)$ ($G$
is the Gauss-Bonnet term) \cite{nojiri2005modified} and $f(G,T)$
\cite{doi:10.1142/S021827181850044X,yousaf2018structure,yousaf2019role,shamir2019stellar}
etc. (details of MTG can be seen in,~\cite{bamba2013modified,
yousaf2016causes,yousaf2016influence,NOJIRI20171,nojiri2008dark}).
The considerable amount of research about the introduction as well
the application of Palatini $f(R)$ scheme is available in literature
\cite{bambi2016wormholes,olmo2011palatini,olmo2015nonsingular,doi:10.1142/S0217732319503334}.

Gravitational collapse is one of the interesting phenomena of the
stellar systems in which the matter of the massive structure moves
towards its center as a consequence of a force produced by its own
gravitational pull. Stars, star clusters and galaxies could be the
final fates of this process from the interstellar gas. The
inhomogeneous state of energy density is believed to be the
progenitors of the stellar collapse. Thus, the importance of the
study of inhomogeneous energy density in the study of collapse is
justified. Penrose \cite{penrose1979singularities} was curious to
understand radiating self-gravitating systems from the clumped
matter distribution. He expressed the heterogeneous energy density
through Weyl curvature tensor and described gravitation in terms of
entropy.

Eardley and Smarr \cite{eardley1979time} calculated exact solution
of a non-static spherical dust cloud in order to explore the
importance of the irregular distribution of energy density and
claimed that such a relativistic collapse is likely to end up with a
new type of naked singularity. Herrera et al. \cite{herrera1998role}
used an analytical approach to evaluate the expression of active
gravitational mass supporting the existence of heterogeneous energy
density for the relativistic radiating spheres. Herrera et al.
\cite{herrera2004spherically} orthogonal splitting of Riemann tensor
and found few scalars associated supporting the existence of
heterogeneous energy density for the relativistic radiating spheres.
Bamba et al. \cite{BAMBA2011451} calculated the rate of spherical
collapse in MTG and discussed some various characteristics of
curvature singularity during its collapse. Yousaf et al. studied the
role of tilted congruences
\cite{PhysRevD.95.024024,doi:10.1139/cjp-2017-0214} as well as MTG
on the existence of regular energy density
\cite{yousaf2018some,RYousaf2019}, and pace of gravitational
collapse \cite{bhatti2020stability, yousaf2020construction,
yousaf2020gravastars}. They inferred that extra curvature terms due
to MTG tend to slow the collapse process.

In the continuation of the study of energy density irregularities,
Herrera \cite{PhysRevD.97.044010} introduced a new concept for the
study of homogeneous distribution of static systems with the help of
a factor. He called this factor as a complexity factor (CF) and
expressed it through the structure scalars. Abbas and Nazar
\cite{Abbas2018510} applied the same procedure in order to present
the definition CF for the anisotropic system in a particular MTG
gravity.  Then, Herrera et al. \cite{PhysRevD.98.104059} generalized
their concept for dynamical self-gravitating systems. Recently,
Sharif and Majid \cite{SHARIF201938} and Yousaf \emph{et al.}
\cite{tayyab1,tayyab2} modified their results in the Brans-Dicke
theory and $f(R,T,R_{\mu\nu}T^{\mu\nu})$ gravity, respectively and
analyzed the role of modified terms in the formulation of CF.

We are adopting the definition of CF mentioned in
\cite{PhysRevD.97.044010}, generalizing it to the static spherically
symmetric case, within the context of a modified gravity theory (the
Palatini $f(R)$ theory). This work is devoted to understanding the
role of Palatini $f(R)$ terms in the theoretical modeling of static
self-gravitating systems. In this regard, we shall present CF and
then this function in terms of structure scalars. The paper is
outlined below. In the coming section, we shall compute Palatini
$f(R)$ equations of motion and hydrostatic equation. After
calculating the Misner-Sharp mass function in the same section, we
shall express the Tolman mass through usual and effective matter
variables. Section 3 is devoted to the orthogonal splitting of the
curvature tensor. The scalars obtained from this calculation will
then be used to define CF. Finally, we present concluding remarks.

\section{Palatini $f(R)$ Gravity and its Related Variables}

The action function for the evaluation of field equation in $f(R)$ gravity is
\begin{equation}\label{1}
S_{f(R)}=\frac{1}{2\kappa}\int d^4x\sqrt{-g}f(\hat{R})+S_M,
\end{equation}
in which $S_M$ describes matter action, while $\kappa$ is a
coupling constant. It is the worthy to stress that in the above
action the scalar function $\hat{R}$ is assembled from the
contraction of the corresponding metric tensor with that of Ricci
tensor associated with the connection symbol ($\hat{R} :=
g^{\gamma\delta}R_{\gamma\delta}$), thereby indicating $\hat{R}$
through geometrical connections. The details on this account can be
found from \cite{PhysRevD.86.127504}. By preserving the relation
$\Gamma_{~\gamma\delta}^\mu\neq\Gamma_{~\delta\gamma}^\mu$, the
variations of Eq.\eqref{1} with $g_{\gamma\delta}$ and
$\Gamma^\mu_{\gamma\delta}$, respectively give
\begin{eqnarray}\label{2}
&&f_R(\hat{R}){\hat{R}}_{\gamma\delta}-[g_{\gamma\delta}f(\hat{R})]/2
={\kappa}T_{\gamma\delta},\\\label{3} &&
\hat{\nabla}_\mu(g^{\gamma\delta}\sqrt{-g}f_R(\hat{R}))=0,
\end{eqnarray}
where $T_{\gamma\delta}$ stands for energy-momentum tensor that here does not depend on geometric connections. Its value can be found as
\begin{align}\label{2a}
T_{\gamma\delta}=-2(-g)^{-1/2}\frac{\delta S_M}{\delta g^{\gamma\delta}}.
\end{align}
The aim of the paper is to calculate CF involved in the emergence of
irregularities over the anisotropic matter distribution in Palatini $f(R)$ gravity. Therefore,
we assume energy momentum in mixed form ($T^{\mu}_{\nu}$) as
\begin{eqnarray}\label{4}
T^{\mu(m)}_{\nu} &=& \rho u^{\mu}u_{\nu}+\Pi
^{\mu}_{\nu}-Ph^{\mu}_{\nu},
\end{eqnarray}
where $u_\gamma$ is the fluid's four velocity and $h^{\mu}_{\nu}$ is
the projection tensor. Further, $P$ is expressed as the combination
of radial $P_r$ and tangential $P_\bot$ pressure components as
$P=\frac{1}{3}(P_{r}+2P_{\bot})$, while the $\Pi^{\mu}_{\nu}$ is the
anisotropic tensor which can be written through anisotropic scalar
$\Pi =-(P_{\bot}-P_{r})$ and four vector $\chi^{\mu}$ as
$\Pi^{\mu}_{\nu}=\Pi(\chi^{\mu}\chi_{\nu}+\frac{1}{3}h^{\mu}_{\nu})$.

The application of $df/d\hat{R}$ has been described through the
subscript $R$ on the associated mathematical quantities. The
dependence of the Ricci invariant on $T$ with
$T:=g^{\gamma\delta}T_{\gamma\delta}$ can be obtained from Eq.\eqref{2}
as
\begin{equation}\label{5}
\hat{R}f_R(\hat{R})-2f(\hat{R})={\kappa}T.
\end{equation}
In the background of vacuum space, $R$ will have a constant value
that can be described through $f(\hat{R})$. This makes us to define
another form of metric tensor $h_{\gamma\delta}$ as
$h_{\gamma\delta}:=f_Rg_{\gamma\delta}$. This scenario describes the
gravitational interaction in the $\Lambda$ dominated epoch. In this
way, the cosmic features of vacuum space from $f(R)$ theory of
gravity can be viewed. Thus, the geometric connection associated
with $h_{\gamma\delta}$ becomes connection of Levi-Civita given by
 \begin{equation}\label{5a}
\Gamma^\nu_{\gamma\delta}=\frac{1}{2}h^{\nu\alpha}(\partial_\gamma h_{\alpha\delta}+\partial_\delta h_{\alpha\gamma}-\partial_\alpha h_{\gamma\delta}).
\end{equation}
There is a conformal relation between tensors $h_{\gamma\delta}$ and
$g_{\gamma\delta}$. In order to proceed our analysis with the second
order metric equations, we compute connection from Eq.(\ref{2}),
which after using in Eq.(\ref{3}) provides
\begin{eqnarray}\nonumber
&&\frac{1}{f_R}\left(\hat{\nabla}_\gamma\hat{\nabla}_\delta-g_{\gamma\delta}
\hat{\Box}\right)f_R+\frac{1}{2}g_{\gamma\delta}\hat{R}+\frac{\kappa}{f_R}
T_{\gamma\delta}+\frac{1}{2}g_{\gamma\delta}\left(\frac{f}{f_R}-\hat{R}\right)
\\\label{4}
&&+\frac{3}{2f_R^2}\left[\frac{1}{2}g_{\gamma\delta}(\hat{\nabla}
f_R)^2-\hat{\nabla}_\gamma f_R\hat{\nabla}_\delta
f_R\right]-\hat{R}_{\gamma\delta}=0,
\end{eqnarray}
while $\hat{\nabla}_\gamma$ indicates covariant derivation with respect to $g_{\gamma\delta}$, $\hat{\Box}$ is the d'Alembertian and
$f_{\mathcal{P}}$=$\frac{df(\mathcal{P})}{d\mathcal{P}}$. Equation \eqref{4} is a single set of field equation that can be interpreted
through Einstein tensor ($G_{\gamma\delta}$) as
\begin{equation}\label{6}
\hat{G}_{\gamma\delta}=\frac{\kappa}{f_R}(T_{\gamma\delta}
+{\mathcal{T}_{\gamma\delta}}),
\end{equation}
where
\begin{eqnarray*}\nonumber
{\mathcal{T}_{\gamma\delta}}&=&\frac{1}{\kappa}\left(\hat{\nabla}_\gamma\hat{\nabla}_
\delta-g_{\gamma\delta}\hat{\Box}\right)f_R-\frac{f_R}{2\kappa}g_{\gamma\delta}
\left(\hat{R}-\frac{f}{f_R}\right)\\\nonumber
&+&\frac{3}{2{\kappa}f_R}\left[\frac{1}{2}g_{\gamma\delta}(\hat{\nabla}
f_R)^2-\hat{\nabla}_\gamma f_R\hat{\nabla}_\delta f_R\right].
\end{eqnarray*}

The system under consideration consists of a static spherically symmetric spacetime whose line element can be given as follows
\begin{equation}\label{9}
ds^{2}=e^{\upsilon}dt^{2}-e^{\omega}dr^{2}-r^{2}(d\theta^{2}
+\sin^{2}\theta d\phi^{2}),
\end{equation}
where $\upsilon$ and $\omega$ are functions of radial coordinate only. The derivation with respect to $r$ will be described through prime in this paper. Under comoving reference system, we define four vectors through metric coefficients as
\begin{equation}\label{7}
\quad \chi^{\mu} =
\bigg(0,{e^{\frac{-\omega}{2}}},0,0\bigg),
\end{equation}
obeying
\begin{equation}\label{8}
\quad \chi^{\gamma}\chi_{\gamma}=-1, \quad
\chi^{\gamma}u_{\gamma}=0.
\end{equation}
The non-zero $f(R)$ equations of motion \eqref{6} for the metric
\eqref{9} are
\begin{align}\nonumber
&\frac{1}{f_R}\left[8\pi\mu+\bigg(\frac{f-Rf_R}{2}\bigg)-\frac{3f_R^{'2}}{4f_Re^{\omega}}-\frac{\upsilon{'}f_R'}{2e^{\omega}}
-\frac{f_R^{'2}}{e^{\omega}}+\frac{f_R{''}}{e^{\omega}}\right]\\\label{f1}
&=\frac{1}{r^{2}}+e^{-\omega}\bigg(\frac
{\omega'}{r}-\frac{1}{r^{2}}\bigg),\\\nonumber
&\frac{1}{f_R}\left[8\pi
P_{r}+\bigg(\frac{f-Rf_R}{2}\bigg)+\frac{5f_R^{'2}}{4f_Re^{\omega}}+\frac{f_R{'}\omega{'}}{2e^{\omega}}\right]\\\label{f2}
&=
\frac{1}{r^{2}}+e^{-\omega}\bigg(\frac{\upsilon^{'}}{r}+\frac{1}{r^{2}}
\bigg),\\\nonumber &\frac{1}{f_R}\left[32\pi
P_{\bot}+4\bigg(\frac{f-Rf_R}{2}\bigg)-\frac{3f_R{'}}{4f_Re^{\omega}}-\frac{f_R'}{re^{\omega}}
-\frac{f_R'}{e^{\omega}}+\frac{f_R{''}}{e^{\omega}}\right]\\\label{f3}
&=-{e^{-\omega}}
\bigg\{\omega^{'}\upsilon^{'}-2\upsilon^{''}+\frac{2\omega'}{r}-\upsilon^{'2}
-\frac{2\upsilon'}{r}\bigg\}.
\end{align}
Here we take $\kappa=8\pi$.

\section{Hydrostatic Equation and Mass Functions}

In this section, we shall compute mass
functions described by Misner-Sharp \cite{PhysRev.136.B571} and
Tolman \cite{PhysRev.35.875}. Later on, we express these functions in
terms of matter and dark source variables of the spherically
symmetric spacetime. One can calculate the conservation laws from
Bianchi identities with respect to usual and effective stress-energy
tensors as
\begin{equation}\label{14}
P'_{r}=\frac{2}{r}(P_{\bot}
-P_{r})-\frac{\beta'}{2}(\rho+P_{r}-T^{1(\mathcal{P})}_{1}+T^{0(\mathcal{P})}
_{0})+\frac{2}{r}(T^{1(\mathcal{P})}_{1}-T^{2(\mathcal{P})}_{2})+D_0,
\end{equation}
where
\begin{align}
D_0=\left(-\frac{f'_R}{2f_R}-\frac{\omega'}{2}\right)\left(T^{1(\mathcal{P})}_{1}-T^{0(\mathcal{P})}_{0}\right)+\left(\frac{2}{r}+\frac{f'_R}{f_R}\right)
\left(T^{1(\mathcal{P})}_{1}-T^{2(\mathcal{P})}_{2}\right)
+(T^{1(\mathcal{P})}_{1})_{,1}.
\end{align}
The total quantity of relativistic matter content for the spherical geometry can be calculated through the formalism
provided by Misner-Sharp as \cite{PhysRev.136.B571}
\begin{equation}\label{15}
m(r)=\frac{r}{2}(1-e^{-\omega}),
\end{equation}
which can be reexpressed after making use of Eq.(\ref{f1}) as
\begin{equation}\label{16}
m(r)=4\pi\int^{r}_{0} \frac{r^{2}}{f_R}(\rho+T^{0(\mathcal{P})}_{0})
dr.
\end{equation}
Using Eqs.(\ref{f1})-(\ref{f3}) and Eq.(\ref{15}), it follows that
\begin{align}\nonumber
m&=\frac{4\pi}{3f_R}r^{3}(\rho-P_{r}-T^{1(\mathcal{P})}_{1}+P_{\bot}+T^{0(\mathcal{P})}_{0}
+T^{2(\mathcal{P})}_{2})\\\label{17}
&-\frac{r^{3}}{3}\bigg[\frac{1}{4}e^{-\omega}\bigg(\upsilon^{''}
+\frac{\upsilon^{'2}}{2}+\frac{\omega^{'}}{r}
+\frac{2}{r^{2}}-\frac{\omega^{'}\upsilon^{'}}{2}
-\frac{\upsilon^{'}}{r}-\frac{2e^{\omega}}{r^{2}}\bigg)\bigg].
\end{align}

The well-known Weyl curvature tensor can be written through its electric part $(E^{\sigma\chi})$ and
Levi-Civita tensor ($\eta_{\mu\nu\gamma\sigma}$) as
\begin{equation}\label{21}
C_{\mu\nu\xi\lambda} =
(g_{\mu\nu\gamma\sigma}g_{\xi\lambda\tau\chi}-\eta_{\mu\nu\gamma\sigma}
\eta_{\xi\lambda\tau\chi})u^{\gamma}u^{\tau}E^{\sigma\chi},\quad
\gamma,\sigma,\tau,
\end{equation}
where $g_{\mu\nu\gamma\sigma}:=g_{\mu\gamma}
g_{\nu\sigma}-g_{\mu\sigma}g_{\nu\gamma}$.
One can write $E^{\sigma\chi}$ in another way as
\begin{equation}\label{20}
E_{\mu\nu} = C_{\mu\xi\nu\lambda}u^{\xi}u^{\lambda},
\end{equation}
whose value with the help of Weyl scalar turns out to be
\begin{equation}\label{22}
E^{\sigma\chi} =
\mathcal{E}\left(\frac{1}{3}h^{\sigma\chi}+\chi^{\sigma}\chi^{\chi}\right).
\end{equation}
Its value through structural variables is found as follows
\begin{equation}\label{23}
\mathcal{E}=
-\frac{1}{4}e^{-\omega}\bigg[\upsilon^{''}+\frac{\upsilon^{'2}}{2}
+\frac{\omega^{'}}{r} +\frac{2}{r^{2}}
-\frac{\omega^{'}\upsilon^{'}}{2}
-\frac{\upsilon^{'}}{r}-\frac{2e^{\omega}}{r^{2}}\bigg].
\end{equation}
The expression mentioning in Eq.(\ref{15}) can be manipulated
through this scalar as
\begin{equation}\label{24}
m=\frac{4\pi}{3f_R}r^{3}(\rho-P_{r}-T^{1(\mathcal{P})}_{1}+P_{\bot}+T^{0(\mathcal{P})}_{0}+T^{2(\mathcal{P})}_{2})+\frac{1}{3}r^{3}\mathcal{E},
\end{equation}
from which $\mathcal{E}$ becomes
\begin{align}\label{her29}
\mathcal{E}&=\frac{4\pi}{f_R}(P_{r}-T^{2(\mathcal{P})}_{2}-P_{\bot}+T^{1(\mathcal{P})}_{1})
-\frac{4\pi}{r^{3}}\int^{r}_{0}
{r}^{3}\left(\frac{\rho}{f_R}\right)'d{r}-\frac{4\pi}{r^{3}}\int^{r}_{0}
{r}^{3}\left(\frac{T^{0(\mathcal{P})}_{0}}{f_R}\right)'d{r}.
\end{align}
This expression has been expressed through matter variables and Palatini $f(R)$ corrections.
Such a relation could be helpful to analyze the heterogenous state of energy density. This equation could provide
information about the existence of regular energy density during the subsequent evolution of the spherical static cloud.
Equation (\ref{24}) along with the above equation gives
\begin{equation}\label{her30}
m=\frac{4\pi
r^{3}}{3\tilde{f_R}}(\rho+T^{0(\mathcal{P})}_{0})-\frac{4\pi}{3\tilde{f_R}}\int^{r}_{0}{r}^{3}
(\rho+T^{0(\mathcal{P})}_{0})^{'}d{r},
\end{equation}
where tilde states that the quantity is being evaluated at the present Ricci scalar condition. This relation has directly related mass function, effective form of regular spherical energy density and the role of
heterogeneous energy density on the stability of relativistic spherical interiors.

The value of one of the metric coefficients found in Eq.\eqref{9}
can be calculated from the field equation and  Eq.(\ref{15}) as
\begin{equation}\label{her10}
\beta' = 2 \quad \frac{mf_R+4\pi r^{3}
(P_{r}+T^{1(\mathcal{P})}_{1})}{r(r-2m)}.
\end{equation}
We now use Eq.(\ref{her10}) in Eq.(\ref{14}) to obtain
Tolman-Opphenheimer-Volkoff (TOV) equation in the presence of
Palatini $f(R)$ corrections as
\begin{eqnarray}\nonumber
P'_{r}&=&T^{1(\mathcal{P})}_{1,1}- \frac{mf_R+4\pi
r^{3}(P_{r}+T^{1(\mathcal{P})}_{1})}{r(r-2m)}(\rho+P_{r}+T^{0(\mathcal{P})}_{0}
-T^{1(\mathcal{P})}_{1})\\\label{her11}&+&
\frac{2}{r}(P_{\bot}-P_{r}+T^{1(\mathcal{P})}_{1}-T^{2(\mathcal{P})}_{2}).
\end{eqnarray}

We now suppose that our system is comprised of boundary surface at $r=r_\Omega$, denoted by $\Omega$, which has demarcated our manifolds
into two parts, the exterior and the interior one. The exterior region can be described through the following spacetime
\begin{equation}\label{her19}
ds^2=\bigg(1-\frac{2\mathcal{M}}{r}\bigg)dt^2-\frac{dr^2}{(1-\frac{2\mathcal{M}}{r})}-r^2(d\theta^2+sin^2\theta
d\phi^2),
\end{equation}
while the interior to $\Omega$ is described by Eq.\eqref{9}. In
Eq.\eqref{her19}, $\mathcal{M}$ describes the mass of the
gravitating source. The fundamental forms of Darmois junction
conditions \cite{darmois1927memorial} at $r=r_\Omega$ provides the
smooth matching of \eqref{9} and \eqref{her19} manifolds. These give
\begin{equation}\nonumber
e^{\upsilon_\Omega}= 1-\frac{2\mathcal{M}}{r_{\Omega}},\quad e^{\omega_\Omega}=
(1-\frac{2\mathcal{M}}{r_{\Omega}})^{-1}, \quad [P_r]_\Omega= 0.
\end{equation}

Herrera and Santos \cite{herrera1997local} described the formula
provided by Tolman \cite{PhysRev.35.875} in order to analyze the
amount of matter content within the spherical geometric
distribution. Further, Herrera et al. \cite{herrera1998role}
expressed this relation in terms of structural properties of
self-gravitating system. They also checked its role in the
maintenance of homogeneous energy density over the boundary. One can
write Tolman formula with Palatini $f(R)$ terms as
\begin{equation}\label{29}
m_{T}=4\pi \int^{r_{\Omega}}_{0}
\frac{r^{2}}{f_R}e^{\frac{\upsilon+\omega}{2}}(\rho+P_{r}+2P_{\bot}+T^{0(\mathcal{P})}_{0}+T^{1(\mathcal{P})}_{1}+2T^{2(\mathcal{P})}_{2})dr.
\end{equation}
For the bounded system, it follows from the above equation that
\begin{equation}\label{30}
m_{T}=4\pi \int^{r}_{0}
\frac{r^{2}}{f_R}e^{\frac{\upsilon+\omega}{2}}(\rho+P_{r}+2P_{\bot}+T^{0(\mathcal{P})}_{0}+T^{1(\mathcal{P})}_{1}+2T^{2(\mathcal{P})}_{2}).
\end{equation}
It could be regarded as the corresponding active gravitational mass.
By making use of Eqs.(\ref{f1})-(\ref{f3}) in Eq.(\ref{30}), we have
\begin{equation}\label{31}
m_{T}=\frac{r^{2}}{2\tilde{f_R}}e^{\frac{\upsilon-\omega}{2}}\upsilon^{'}-\frac{4\pi}{\tilde{f_R}}\int
^{r}_{0}
e^{\frac{\upsilon+\omega}{2}}{{r}^{2}}(2T^{2(\mathcal{P})}_{2}+T^{1(\mathcal
{G})}_{1}-T^{0(\mathcal{P})}_{0})d{r}.
\end{equation}
Using Eq.(\ref{her10}) in Eq.\eqref{31}, we get
\begin{eqnarray}\nonumber
m_{T}&=&\bigg(4\pi{r^{3}}(P_{r}+T^{1(\mathcal{P})}_{1})+m\tilde{f_R}\bigg)\frac{e^{\frac{\upsilon
+\omega}{2}}}{\tilde{f_R}}\\\label{32}&-&\frac{4\pi}{\tilde{f_R}}\int^{r}_{0}e^{\frac{\upsilon+\omega}{2}}{{r}
^{2}}(2T^{2(\mathcal{P})}_{2}+T^{1(\mathcal{P})}_{1}-T^{0(\mathcal{P})}_{0})d{r}.
\end{eqnarray}
Then the corresponding $m_{T}$ for the anisotropic system turns out to be
\begin{eqnarray}\nonumber
m_{T}&=&\bigg(\frac{r}{r_{\Omega}}\bigg)^{3}[m_{T}]_{\Omega}+r^{3}
\int^{r_{\Omega}}_{r}\frac{e^{\frac{\upsilon+\omega}{2}}}{{r}}\bigg[\frac{4\pi}{\tilde{f_R}}
(P_{\bot}-P_{r})-\mathcal{E}\bigg]d{r}\\\nonumber&
+&\frac{4\pi}{\tilde{f_R}}r^{3}\int^{r_{\Omega}}_{r}e^{\frac{\upsilon+\omega}{2}}(T^
{0(\mathcal{P})}_{0}+4T^{1(\mathcal{P})}_{1}+2T^{2(\mathcal{P})}_{2})
d{r}\\\label{33}&-&r^3\frac{4\pi}{\tilde{f_R}}\int^{r}_{0}e^{\frac{\upsilon+\omega}
{2}}{{r}^{2}}(T^{0(\mathcal{P})}_{0}-T^{1(\mathcal{P})}_{1}
-2T^{2(\mathcal{P})}_{2})d{r}.
\end{eqnarray}
This equation states that the value of Tolman mass depends mainly on
the participation of fluid energy density, $f(R)$ dark
source terms and pressure anisotropy of the spherical relativistic
geometric distribution.

\section{Structure Scalars}

The definition for Riemann tensor can be written as follows
\begin{equation}\label{18}
R^\mu_{\xi\nu\lambda}=C^\mu_{\xi\nu\lambda}+\frac{1}{2}R^{\mu}_{\nu}g_{\xi\lambda}
-\frac{1}{2}R_{\xi\nu}\delta^{\mu}_{\lambda}+\frac{1}{2}R_{\xi\lambda}\delta^{\mu}_{\nu}-
\frac{1}{2}R^{\mu}_{\lambda}g_{\xi\nu}-\frac{1}{6}R\bigg({\delta^\mu_\nu}g_{\xi\lambda}
-{\delta^\mu_\lambda}g_{\xi\nu}\bigg),
\end{equation}
where $R,~C_{\mu\xi\nu\lambda}$ and $R_{\mu\nu}$ stand for the Ricci scalar, the Weyl tensor  and the Ricci tensor. One can manipulate Eq.\eqref{18} as
\begin{equation}\label{19}
R^{\mu\gamma}_{\nu\delta}=C^{\mu\gamma}_{\nu\delta}
+2T^{(tot)[\mu}_{[\nu}\delta^{\gamma]}_{\delta]}+T^{(tot)}\bigg(\frac{1}{3}
\delta^\mu_{[\nu}\delta^\gamma_{\delta]}-\delta^{[\mu}_{[\nu}\delta^{\gamma]}_{\delta]}\bigg).
\end{equation}
Herrera et al. \cite{herrera2009structure} calculated some scalar
variables from the Riemann tensor splitting. This technique has been
proved to be very helpful in understanding the basic ingredients of
matter content of the self-gravitating system. For this purpose, we
took few tensorial quantities as follows \cite{herrera2009structure}
\begin{eqnarray}\label{34}
Y_{\mu\nu}&=&R_{\mu\xi\nu\lambda}u^{\xi}u^{\lambda},
\\\label{35}
X_{\mu\nu}
&=&^{*}R^{*}_{\mu\xi\nu\lambda}u^{\xi}u^{\lambda}=\frac{1}{2}\eta
^{\alpha\beta}_{\mu\nu}R^{*}_{\alpha\beta\xi\lambda}u^{\xi}u^{\lambda},
\end{eqnarray}
where $R^{*}_{\mu\nu\xi\lambda}=\frac{1}{2}\eta
_{\alpha\beta\xi\lambda}R^{\alpha\beta}_{\mu\nu}$. Equation (\ref{19}) can be written
alternatively as follows
\begin{equation}\nonumber
R^{\mu\xi}_{\nu\lambda}=R^{\mu\xi}_{(I)\nu\lambda}+R^{\mu\xi}_{(II)\nu\lambda}
+R^{\mu\xi}_{(III)\nu\lambda}+R^{\mu\xi}_{(IV)\nu\lambda}+R^{\mu\xi}_{(V)\nu\lambda},
\end{equation}
where
\begin{eqnarray}\nonumber
R^{\mu\xi}_{(I)\nu\lambda}&=&\frac{16\pi}{\tilde{f_R}}\rho{u^{[\mu}u_{[\nu}\delta^{\xi]}_{\lambda]}}
+\frac{8\pi}{\tilde{f_R}}{(\rho-3P)}\bigg(\frac{1}{3}\delta^\mu_{[\nu}\delta^\xi_{\lambda]}
-\delta^{[\mu}_{[\nu}\delta^{\xi]}_{\lambda]}\bigg)
\\\label{36}&-&16\pi{P}{h^{[\mu}_{[\nu}\delta^{\xi]}_{\lambda]}},\\\label{37}
R^{\mu\xi}_{(II)\nu\lambda}&=&\frac{16\pi}{\tilde{f_R}}{\Pi^{[\mu}_{[\nu}\delta^{\xi]}_{\lambda]}},\\\label{38}
R^{\mu\xi}_{(III)\nu\lambda}&=&4u^{[\mu}u_{[\nu}E^{\xi]}_{\lambda]}
-\epsilon^{\mu\xi}_{\alpha}\epsilon_{\nu\lambda\beta}E^{\alpha\beta},
\\\nonumber R^{\mu\xi}_{(IV)\nu\lambda}&=&4\bigg[R_{\gamma\eta}
\delta^{[\mu}_{[\nu}\delta^{\xi]}_{\lambda]}
+g_{\gamma\eta}\delta^{[\mu}_{[\nu}\delta^{\xi]}_{\lambda]}
-R^{[\mu}_{\gamma\eta[\nu}\delta^{\xi]}_{\lambda]}
-R_{\gamma[\nu}\delta^{[\mu}_{\eta}\delta^{\xi]}_{\lambda]}
-R^{[\mu}_\eta g_{\gamma[\nu}\delta^{\xi]}_{\lambda]}
\\\label{39}&-&\frac{1}{2}R\bigg(\delta^{[\mu}_{[\nu}g_{\gamma\eta}\delta^{\xi]}_{\lambda]}
+\delta^{[\mu}_\eta
g_{\gamma[\nu}\delta^{\xi]}_{\lambda]}\bigg)\bigg]
\nabla^\gamma\nabla^\eta f_R
+(f-Rf_R)\delta^{[\mu}_{[\nu}\delta^{\xi]}_{\lambda]},\\\label{40}
R^{\mu\xi}_{(V)\nu\lambda}&=&4[(R_{\gamma\eta}-\frac{1}{2}Rg_{\gamma\eta})
\nabla^\gamma\nabla^\eta
f_R+(f-Rf_R)](\frac{1}{3}\delta^\mu_{[\nu}\delta^\xi_{\lambda]}
-\delta^{[\mu}_{[\nu}\delta^{\xi]}_{\lambda]}),
\end{eqnarray}
and $\epsilon_{\nu\lambda\beta}=u^\mu\eta_{\mu\nu\lambda\beta}$ with
$\epsilon_{\nu\lambda\beta}u^\beta=0$. By making use of Eqs.(\ref{36})-(\ref{40}) in Eqs.(\ref{34}) and (\ref{35}), we get
\begin{eqnarray}\label{41}
X^{(tot)}_{\mu\nu}&=&X^{(m)}_{\mu\nu}+X^{(\mathcal{P})}_{\mu\nu},
\\\label{42} Y^{(tot)}_{\mu\nu}&=&Y^{(m)}_{\mu\nu}+Y^{(\mathcal{P})}_{\mu\nu},
\end{eqnarray}
where the terms with superscripts $m$ and $\mathcal{P}$ indicates
that the corresponding terms are related to usual matter and
Palatini $f(R)$ terms, respectively. These may be written as
\begin{align}\label{43}
X^{(tot)}_{\mu\nu}&=\frac{8\pi}{3\tilde{f_R}}\rho
h_{\mu\nu}+\frac{4\pi}{\tilde{f_R}}\Pi_{\mu\nu}-E_{\mu\nu}+X^{(\mathcal{P})}_{\mu\nu},\\\label{44}
Y^{(tot)}_{\mu\nu}&=\frac{4\pi}{3\tilde{f_R}}(\rho+3P)
h_{\mu\nu}+\frac{4\pi}{\tilde{f_R}}\Pi_{\mu\nu}+E_{\mu\nu},
\end{align}
where $Y^{(tot)}_{\mu\nu}=Y^{(m)}_{\mu\nu}$ \cite{PhysRevD.97.044010}.
One can write both trace and trace-less components from the above equations as
\begin{eqnarray}\label{45}
X_{\mu\nu}&=&X_{TF}\bigg(\frac{1}{3}h_{\mu\nu}+s_{\mu}s_{\nu}\bigg)
+\frac{1}{3}X_{T}h_{\mu\nu},
\\\label{46}
Y_{\mu\nu}&=&Y_{TF}\bigg(\frac{1}{3}h_{\mu\nu}+s_{\mu}s_{\nu}\bigg)
+\frac{1}{3}Y_{T}h_{\mu\nu},
\end{eqnarray}
from where $Y_T$ and $Y_{TF}$ after using the Palatini $f(R)$ field equations can be found as follows
\begin{eqnarray}\label{47}
Y_{T}&=&\frac{4\pi}{\tilde{f_R}}(\rho-2\Pi+3P_{r}), \\\label{48}
Y_{TF}&=&\frac{4\pi}{\tilde{f_R}}\Pi+\mathcal{E}.
\end{eqnarray}
Equations (\ref{her29}) and (\ref{48}) yield
\begin{equation}\label{49}
Y_{TF}=\frac{8\pi}{\tilde{f_R}}\Pi+\frac{4\pi}{\tilde{f_R}}(T^{1(\mathcal{P})}_{1}-T^{2(\mathcal{P})}_{2})-\frac{4\pi}{r^{3}}\int^{r}_{0}
{r}^{3}\left(\frac{\rho}{f_R}\right)'d{r}-\frac{4\pi}{r^{3}}\int^{r}_{0}
{r}^{3}\left(\frac{T^{0(\mathcal{P})}_{0}}{f_R}\right)'d{r}.
\end{equation}
These scalar quantities were defined and discussed in detail for the
first time by Herrera \emph{et al.} \cite{herrera2009structure}. The
expression of $m_T$ can be written after using Eqs.(\ref{48}) and
(\ref{33}) as
\begin{eqnarray}\nonumber
m_{T}&=&\bigg(\frac{r}{r_{\Omega}}\bigg)^{3}[m_{T}]_{\Omega}+r^{3}
\int^{r_{\Omega}}_{r}\frac{e^{\frac{\upsilon+\omega}{2}}}{{r}}Y_{TF}d{r}\\\nonumber&
+&\frac{4\pi}{\tilde{f_R}}r^{3}\int^{r_{\Omega}}_{r}e^{\frac{\upsilon+\omega}{2}}(T^
{0(\mathcal{P})}_{0}+4T^{1(\mathcal{P})}_{1}+2T^{2(\mathcal{P})}_{2})
d{r}\\\label{50}&-&r^3\frac{4\pi}{\tilde{f_R}}\int^{r}_{0}e^{\frac{\upsilon+\omega}
{2}}{{r}^{2}}(T^{0(\mathcal{P})}_{0}-T^{1(\mathcal{P})}_{1}
-2T^{2(\mathcal{P})}_{2})d{r}.
\end{eqnarray}
The comparison of Eq.(\ref{33}) and Eq.(\ref{50}) provides
\begin{eqnarray}\label{51}
\int^{r_{\Omega}}_{r}\frac{e^{\frac{\beta+\alpha}{2}}}{{r}}
Y_{TF}d{r}&=&\int^{r_{\Omega}}_{r}\frac{e^{\frac{\beta+\alpha}
{2}}}{{r}}[\frac{4\pi}{\tilde{f_R}}(P_{r}-P_{\bot})+\mathcal{E}]d{r},
\end{eqnarray}
which implies that $Y_{TF}$ is correlated with the fluid
distribution's influence of irregular energy density, $f(R)$
corrections, and anisotropic pressure.

\section{The Complexity Factor}

Due to some physical quantities, i.e., inhomogeneity of mass
distribution of spherical structure, heat radiations and fluid's
viscosity, etc, a CF of a corresponding system can be generated. The
simplest system with null CF can be considered normally as a system
coupled with an isotropic pressure and regular energy density. The
origins of CF here could be anisotropic pressure, MTG terms of
Palatini $f(R)$ gravity along with irregular density in the
relativistic matter content. One of the scalars, $Y_{TF}$ appearing
in Eq.(\ref{49}) can be called as CF in our case as it appears in
the expression of Tolman mass. Our system of partial differential
equations includes unknown structural variables. In order to treat
such scenario, we restrict our system to enter in the less-complex
state. This could be possible by using $Y_{TF}=0$ in the analysis.
Under this background, we have
\begin{equation}\label{52}
\Pi=\frac{\tilde{f_R}}{2r^{3}}\int^{r}_{0}
{r}^{3}\left(\frac{\rho}{f_R}\right)'d{r}+\frac{\tilde{f_R}}{2r^{3}}\int^{r}_{0}
{r}^{3}\left(\frac{T^{0(\mathcal{P})}_{0}}{f_R}\right)'d{r}+\frac{1}{2}(T^{1(\mathcal{P})}_{1}-T^{2(\mathcal{P})}_{2}).
\end{equation}
In this way, we have evaluated the value of $\Pi$. Equation
(\ref{52}) describes the modeling of anisotropic spherical system
evolving with zero CF with the extra degrees of freedom mediated by
Palatini $f(R)$ terms. To find rest of unknown, it could be helpful
to consider viable analytical models from the literature. We shall
take a model known widely as Gokhroo-Mehra ansatz (GMA).

Abedi \emph{et al.} \cite{s2} explored the energy-momentum complex
for the flat Friedmann-Robertson-Walker manifold within an
environment of GR and one of the modified gravity models. It is
worthy to stress that our calculated CF could be related to the
energy-momentum complex because it depends on the stress-energy
momentum tensor. This suggests that our approach could be related to
the definition of gravitational stress-energy
pseudo-tensor \cite{s1}.\\

\textbf{The Gokhroo-Mehra Ansatz}\\

The interior solutions for the spherically symmetric anisotropic
fluid configurations having irregular distribution of energy density
were studied by Gokhroo and Mehra \cite{gokhroo1994anisotropic}.
They assumed the following choice of energy density as
\begin{equation}\label{53}
\rho=\rho_{0}\bigg(1-\frac{Kr^{2}}{r^{2}_{\Omega}}\bigg),
\end{equation}
in which they used $\rho_0$ as a constant quantity, along with
$K\in(0,1)$. By making use of  such selection of energy density,
Eq.(\ref{16}) may be written alternatively as
\begin{equation}\label{54}
m(r)=\frac{4\pi}{\tilde{f_R}}\int^{r}_{0}{{r}^{2}}{T^{0(\mathcal{P})}_{0}}d\hat{r}+\frac{4\pi
r^{3}}{3\tilde{f_R}}\rho_{0}\bigg(1-\frac{3Kr^{2}}{5r^{2}_{\Omega}}\bigg),
\end{equation}
which upon using Eq.(\ref{15}) provides
\begin{equation}\label{55}
e^{-\upsilon}=1+\frac{3K\xi_0
r^{4}}{5r^{2}_{\Omega}\tilde{f_R}}-\frac{\xi_0
r^{2}}{\tilde{f_R}}-\frac{8\pi}{r\tilde{f_R}}\int^{r}_{0}{{r}^{2}}{T^{0(\mathcal{P})}_{0}}d{r},
\end{equation}
where $\xi_0=8\pi\rho_{0}/3$. From Eqs.(\ref{f2}) and
(\ref{f3}), we have
\begin{equation}\label{56}
\frac{8\pi}{f_R}(P_{r}-P_{\bot}+T^{1(\mathcal{P})}_{1}-T^{2(\mathcal{P})}_{2})=e^{-\upsilon}\bigg[\frac{\omega^{'}}{2r}
+\frac{1}{r^{2}}+\frac{\upsilon^{'}\omega^{'}}{4}
+\frac{\upsilon^{'}}{2r}-\frac{\omega^{''}}{2}
-\frac{\omega^{'2}}{4}\bigg]-\frac{1}{r^2}.
\end{equation}
It could be fine to use couple of effective form of new variables as
follows
\begin{equation}\label{57}
e^{\omega(r)}=\frac{1}{e^{\int(\frac{2}{r}-2z(r))dr}},\quad
e^{\upsilon}=\frac{1}{y(r)}.
\end{equation}
After using above variables, Eq.(\ref{56}) turns out to be
\begin{equation}\label{58}
y^{'}+\bigg[2z+\frac{2z^{'}}{z}+\frac{4}
{r^{2}z}-\frac{6}{r}\bigg]y=-\bigg[\frac{1}{r^{2}}+\frac{8\pi}{f_R}(\Pi
+{T^{1(\mathcal{P})}_{1}}-{T^{2(\mathcal{P})}_{2}})\bigg]
\frac{2}{z}.
\end{equation}
The spacetime with its metric coefficients through $z$ and $\Pi$ can
be written as \cite{herrera2008all}
\begin{eqnarray}\nonumber
ds^{2}&=&e^{\int(2z(r)-\frac{2}{r})dr}dt^{2}-r^{2}d\theta^{2}
-r^{2}\sin^{2}\theta d\phi^{2}\\\label{59} &+&\frac{z^{2}(r)e^{\int
(2z(r)-\frac{4}{r^{2}z(r)})dr}}{2r^{6}\int \frac{e^{\int
(2z(r)-\frac{4}{r^{2}z(r)})dr}z(r)(\frac{1}{r^2}+8\pi f_R^{-1}(\Pi
+{T^{1(\mathcal{P})}_{1}}-{T^{2(\mathcal{P})}_{2}}))}{r^{6}}dr
+C}dr^{2},
\end{eqnarray}
where $C$ is an integration constant. In this framework, the values of structural variables are
\begin{eqnarray}
\label{60} \frac{4\pi}{f_R}(\rho+{T^{0(\mathcal{P})}_{0}})&=&\frac{m^{'}}{r},
\\\label{61}
\frac{4\pi }{f_R}(P_r+{T^{1(\mathcal{P})}_{1}})&=&\frac{\frac{m}{r}-z(2m-r)-1}{4\pi
r^{2}},\\\label{62}
\frac{8\pi}{f_R}(P_\bot+{{T^{2(\mathcal{P})}_{2}}})&=&z\bigg(\frac{m}{r^{2}}-\frac{m^{'}}{r}\bigg)
+\bigg(z^{2}+z^{'}+\frac{1}{r^{2}}-\frac{z}{r}\bigg)\bigg
(1-\frac{2m}{r}\bigg).
\end{eqnarray}
These equations have formulated the matter variables in terms of
$m,~z$ and $f(R)$ dark source terms. After selecting some suitable
initial condition, one can solve (analytically or numerically) them
for different choices of $z$ and $m$. Then the resultant variables
will be expressed in terms of Palatini $f(R)$ dark source terms.
Such equations objectively analyze the existence of celestial bodies
for $Y_{TF}=0$.

\section{Concluding Remarks}

In this paper, we study the dynamics of a non-rotating spherically
symmetric spacetime which is coupled with a locally anisotropic
fluid distribution in a non-linear $f(R)$ theory of gravity. Such a
mathematical model can be called intuitively as complex one, thereby
indicating a need to explore the corresponding complexity factor.
The purpose of this work is to explore this factor in the context of
$f(R)$ gravity.

\begin{itemize}

\item It is well-known that the static relativistic pressure isotropic
spheres with a regular energy density distribution could be regarded
as the less complex systems. Therefore, it would be justified to
consider null contribution of CF for those
systems. It has been analyzed in GR that one of the structure
scalars $Y_{TF}$ is the CF. The same result is found for Palatini
$f(R)$ gravity with the difference that the dark source terms of
Palatini $f(R)$ terms appearing in the definition of $Y_{TF}$ are
slowing down such transition due to their non-attractive nature.

\item Another very important result stems form the expression of
$Y_{TF}$ is that it contains spherical structural effects coming
from the irregular energy density, Palatini $f(R)$ terms and
effective form of local anisotropic pressure fabricated in a
particular way. This factor has been found to be zero for those
systems who evolves with a homogeneous perfect fluid in GR. But in
our case the extra curvature Palatini $f(R)$ terms are providing
resistance to the system in leaving their homogeneous state. We
expect to see such analysis in the presence of electromagnetic
field.
\item The structure variable $Y_{TF}$ is specifying the role of irregular energy density, Palatini $f(R)$ terms and local anisotropic
terms in a particular order.
\item The same modified scalar variable has been found to be involved in estimating the digression of
the Tolman mass $m_T$ for regular spheres, mediated by irregularity in the energy density of the anisotropic matter
configurations.
\item All of our results reduce to GR \cite{PhysRevD.97.044010} under the condition $f(R)=R$.

\end{itemize}

We have assumed a Palatini $f(R)$ gravity, one can induce further
degrees of freedom in the analysis of relativistic ideal and
non-ideal configurations. Capozziello \emph{et al.} \cite{j1}
associated Gauss-Bonnet curvature terms with the stress energy
tensor of an ideal matter content. They have also studied the
geometric interpretation of the dark components of the cosmological
Hubble flow. As an extension in the gravitational component of GR
action, the further degrees of freedom in this part can be modeled
theoretically as ideal matter configurations sourcing the
corresponding equations of motion \cite{j2}.

\vspace{0.25cm}

{\bf Acknowledgments}

\vspace{0.25cm}

This work has been supported financially by National Research
Project for Universities (NRPU), Higher Education Commission,
Pakistan under research project No. 8754/Punjab/NRPU/R\&D/HEC/2017.

\vspace{0.5cm}

\section*{Appendix}
\begin{eqnarray*}\nonumber
T^{0(\mathcal{P})}_{0}&=&\frac{1}{2}f_{\mathcal{P}}e^{\beta}
+\frac{e^{\beta-2\alpha}}{r^2}\bigg[(3\beta^{'}\alpha{'}
-{\beta^{'}}^{2}-2\beta^{''}+e^{\alpha}({{\beta{'}}^{2}}
+2\beta^{''}-\beta^{'}\alpha^{'}))f_{\mathcal{P}}
\\\label{72}&+&2(e^{\alpha}-3)\alpha^{'}{\mathcal{P}}^{'}
f_{\mathcal{P}\mathcal{P}}+4\mathcal{P}^{''}(1-e^{\alpha}
)f_{\mathcal{P}\mathcal{P}}+4{\mathcal{P}^{'}}^{2}
(1-e^{\alpha})f_{{\mathcal{P}}{\mathcal{P}}{\mathcal{P}}}\bigg],
\\\nonumber
T^{1(\mathcal{P})}_{1}&=&-\frac{1}{2}f_{\mathcal{P}}e^{\alpha}
+\frac{1}{r^{2}}\bigg[\bigg(\beta^{'}\alpha^{'}(1-\frac{3}
{e^{\alpha}})+2{\beta}^{''}(\frac{1}{e^{\alpha}}-1)+{{\beta}^{'}}^{2}
(\frac{1}{e^{\alpha}}-1)\bigg)f_{\mathcal{P}}\\\label{73}&+&\bigg
(2{\beta}^{'}{\mathcal{P}}^{'}(1-\frac{3}{e^{\alpha}})\bigg)
f_{\mathcal{P}\mathcal{P}}\bigg],\\\nonumber
T^{2(\mathcal{P})}_{2}&=&-\frac{1}{2}{r^{2}}f_\mathcal{P}
+\bigg(\frac{{\beta^{'}}^{2}}{e^{2\alpha}
}-\frac{{\beta^{'}}^{2}}{e^{\alpha}}-2\frac{\beta^{''}}
{e^{\alpha}}+2\frac{\beta^{''}}{e^{2\alpha}
}+\frac{\beta^{'}\alpha^{'}}{e^{\alpha}}-3\frac{\beta^{'}
\alpha^{'}}{e^{2\alpha}}\bigg)f_{\mathcal{P}}\\\label{74}&-&\bigg
((r{\beta^{'}}^{2}+2r{\beta}^{''}-3r{\beta^{'}}{\alpha^{'}})
{\mathcal{P}^{'}}+{2r{\beta}^{'}{\mathcal{P}}^{''}}\bigg)\frac
{f_{\mathcal{P}\mathcal{P}}}{e^{2\alpha}}-\frac{2r{\beta^
{'}}{\mathcal{P}^{'}}^{2}}{e^{2\alpha}}f_{\mathcal{P}\mathcal{P}\mathcal{P}}.
\end{eqnarray*}


\vspace{0.5cm}

\begin{thebibliography}{10}

\bibitem{pietrobon2006integrated}
D.~Pietrobon, A.~Balbi, and D.~Marinucci {\em Phys. Rev. D},
vol.~74,
  p.~043524, 2006.

\bibitem{giannantonio2006high}
T.~Giannantonio {\em et~al.} {\em Phys. Rev. D}, vol.~74, p.~063520,
2006.

\bibitem{riess2007new}
A.~G. Riess {\em et~al.} {\em Astrophys. J.}, vol.~659, p.~98, 2007.

\bibitem{ostriker1973numerical}
J.~P. Ostriker and P.~J. Peebles {\em Astrophys. J.}, vol.~186,
p.~467, 1973.

\bibitem{refregier2003weak}
A.~Refregier {\em Annu. Rev. Astron. Astrophys.}, vol.~41, p.~645,
2003.

\bibitem{abbott2017gw170817}
B.~P. Abbott {\em et~al.} {\em Phys. Rev. Lett.}, vol.~119,
p.~161101, 2017.

\bibitem{abbott2016observation}
B.~P. Abbott {\em et~al.} {\em Phys. Rev. Lett.}, vol.~116,
p.~061102, 2016.

\bibitem{abbott2017gw170814}
B.~P. Abbott {\em et~al.} {\em Phys. Rev. Lett.}, vol.~119,
p.~141101, 2017.

\bibitem{laureijs2011euclid}
R.~Laureijs {\em et~al.} {\em arXiv preprint arXiv:1110.3193}, 2011.

\bibitem{lombriser2017challenges}
L.~Lombriser and N.~A. Lima {\em Phys. Lett. B}, vol.~765, p.~382,
2017.

\bibitem{baker2017strong}
T.~Baker, E.~Bellini, P.~G. Ferreira, M.~Lagos, J.~Noller, and
I.~Sawicki {\em
  Phys. Rev. Lett.}, vol.~119, p.~251301, 2017.

\bibitem{senovilla20151965}
J.~M.~M. Senovilla and D.~Garfinkle {\em Class. Quantum Grav.},
vol.~32,
  p.~124008, 2015.

\bibitem{qadir2017modified}
A.~Qadir, H.~W. Lee, and K.~Y. Kim {\em Int. J. Mod. Phys. D},
vol.~26,
  p.~1741001, 2017.

\bibitem{capozziello2010beyond}
S.~Capozziello and V.~Faraoni, {\em Beyond Einstein Gravity},
vol.~170.
\newblock Springer Science \& Business Media, 2010.

\bibitem{capozziello2011extended}
S.~Capozziello and M.~De~Laurentis {\em Phys. Rep.}, vol.~509,
p.~167, 2011.

\bibitem{bamba2012dark}
K.~Bamba, S.~Capozziello, S.~Nojiri, and S.~D. Odintsov {\em
Astrophys. Space
  Sci.}, vol.~342, p.~155, 2012.

\bibitem{doi:10.1142/S021773231950189X}
M.~J. Khan, G.~Shabbir, and M.~Ramzan {\em Mod. Phys. Lett. A},
vol.~34,
  p.~1950189, 2019.

\bibitem{sahu2017cosmic}
S.~K. Sahu, S.~K. Tripathy, P.~K. Sahoo, and A.~Nath {\em Chin. J.
Phys.},
  2017.

\bibitem{nojiri2007introduction}
S.~Nojiri and S.~D. Odintsov {\em Int. J. Geom. Meth. Mod. Phys.},
vol.~4,
  p.~115, 2007.

\bibitem{elizalde2011nonsingular}
E.~Elizalde, S.~Nojiri, S.~D. Odintsov, L.~Sebastiani, and
S.~Zerbini {\em
  Phys. Rev. D}, vol.~83, p.~086006, 2011.

\bibitem{bamba2010cosmological}
K.~Bamba, C.-Q. Geng, and C.-C. Lee {\em J. Cosmol. Astropart.
Phys.},
  vol.~2010, p.~021, 2010.

\bibitem{houndjo2017higher}
M.~J.~S. Houndjo, M.~E. Rodrigues, N.~S. Mazhari, D.~Momeni, and
R.~Myrzakulov
  {\em Int. J. Mod. Phys. D}, vol.~26, p.~1750024, 2017.

\bibitem{doi:10.1142/S0219887818501463}
Z.~Yousaf, M.~Sharif, M.~Ilyas, and M.~Z. Bhatti {\em Int. J. Geom.
Meth. Mod.
  Phys.}, vol.~15, p.~1850146, 2018.

\bibitem{nojiri2005modified}
S.~Nojiri and S.~D. Odintsov {\em Phys. Lett. B}, vol.~631, p.~1,
2005.

\bibitem{doi:10.1142/S021827181850044X}
M.~Z. Bhatti, M.~Sharif, Z.~Yousaf, and M.~Ilyas {\em Int. J. Mod.
Phys. D},
  vol.~27, p.~1850044, 2018.

\bibitem{yousaf2018structure}
Z.~Yousaf {\em Astrophys. Space Sci.}, vol.~363, p.~226, 2018.

\bibitem{yousaf2019role}
Z.~Yousaf {\em Eur. Phys. J. Plus}, vol.~134, p.~245, 2019.

\bibitem{shamir2019stellar}
M.~F. Shamir and M.~Ahmad {\em Mod. Phys. Lett. A}, vol.~34,
p.~1950038, 2019.

\bibitem{bamba2013modified}
K.~Bamba, S.~Nojiri, and S.~D. Odintsov {\em arXiv preprint
arXiv:1302.4831},
  2013.

\bibitem{yousaf2016causes}
Z.~Yousaf, K.~Bamba, and M.~Z. Bhatti {\em Phys. Rev. D}, vol.~93,
p.~124048,
  2016.

\bibitem{yousaf2016influence}
Z.~Yousaf, K.~Bamba, and M.~Z. Bhatti {\em Phys. Rev. D}, vol.~93,
p.~064059,
  2016.

\bibitem{NOJIRI20171}
S.~Nojiri, S.~Odintsov, and V.~Oikonomou {\em Phys. Rep.}, vol.~692,
p.~1,
  2017.

\bibitem{nojiri2008dark}
S.~Nojiri and S.~D. Odintsov {\em arXiv preprint arXiv:0807.0685},
2008.

\bibitem{bambi2016wormholes}
C.~Bambi, A.~Cardenas-Avendano, G.~J. Olmo, and D.~Rubiera-Garcia
{\em Phys.
  Rev. D}, vol.~93, p.~064016, 2016.

\bibitem{olmo2011palatini}
G.~J. Olmo and D.~Rubiera-Garcia {\em Phys. Rev. D}, vol.~84,
p.~124059, 2011.

\bibitem{olmo2015nonsingular}
G.~J. Olmo and D.~Rubiera-Garcia {\em Universe}, vol.~1, p.~173,
2015.

\bibitem{doi:10.1142/S0217732319503334}
Z.~Yousaf {\em Mod. Phys. Lett. A}, vol.~34, p.~1950333, 2019.

\bibitem{penrose1979singularities}
R.~Penrose, ``Singularities and time-asymmetry,'' in {\em General
Relativity:
  An Einstein Centenary Survey} (W.~Israel and S.~W. Hawking, eds.), p.~581,
  Cambridge University Press, 1979.

\bibitem{eardley1979time}
D.~M. Eardley and L.~Smarr {\em Phys. Rev. D}, vol.~19, p.~2239,
1979.

\bibitem{herrera1998role}
L.~Herrera, A.~Di~Prisco, J.~L. Hern{\'a}ndez-Pastora, and N.~O.
Santos {\em
  Phys. Lett. A}, vol.~237, p.~113, 1998.

\bibitem{herrera2004spherically}
L.~Herrera, A.~Di~Prisco, J.~Martin, J.~Ospino, N.~O. Santos, and
O.~Troconis
  {\em Phys. Rev. D}, vol.~69, p.~084026, 2004.

\bibitem{BAMBA2011451}
K.~Bamba, S.~Nojiri, and S.~D. Odintsov {\em Phys. Lett. B},
vol.~698, p.~451,
  2011.

\bibitem{PhysRevD.95.024024}
Z.~Yousaf, K.~Bamba, and M.~Z. Bhatti {\em Phys. Rev. D}, vol.~95,
p.~024024,
  2017.

\bibitem{doi:10.1139/cjp-2017-0214}
Z.~Yousaf, M.~Z. Bhatti, and A.~Rafaqat {\em Can. J. Phys.},
vol.~95, p.~1246,
  2017.

\bibitem{yousaf2018some}
Z.~Yousaf and M.~Z. Bhatti {\em Int. J. Geom. Meth. Mod. Phys.},
vol.~15,
  p.~1850160, 2018.

\bibitem{RYousaf2019}
Z.~Yousaf, M.~Z. Bhatti, and R.~Saleem {\em Eur. Phys. J. Plus},
vol.~134,
  p.~142, 2019.

\bibitem{bhatti2020stability}
M.~Z.~Bhatti, Z.~Yousaf, and M.~Yousaf {\em Phys. Dark Universe},
vol.~28,
  p.~100501, 2020.

\bibitem{yousaf2020construction}
Z.~Yousaf, {\em Phys. Dark Universe}, vol.~28,
  p.~100509, 2020.

\bibitem{yousaf2020gravastars}
Z.~Yousaf, M.~Z.~Bhatti, and H.~Asad {\em Phys. Dark Universe},
vol.~28,
  p.~100527, 2020.

\bibitem{PhysRevD.97.044010}
L.~Herrera {\em Phys. Rev. D}, vol.~97, p.~044010, 2018.

\bibitem{Abbas2018510}
G.~Abbas and H.~Nazar {\em Eur. Phys. J. C}, vol.~78, p.~510, 2018.

\bibitem{PhysRevD.98.104059}
L.~Herrera, A.~Di~Prisco, and J.~Ospino {\em Phys. Rev. D}, vol.~98,
p.~104059,
  2018.

\bibitem{SHARIF201938}
M.~Sharif and A.~Majid {\em Chin. J. Phys.}, vol.~61, p.~38, 2019.

\bibitem{tayyab1}
Z.~Yousaf,~M.~Z.~Bhatti, T.~Naseer and I. Ahmad {\em Phys. Dark
Universe}, vol.~29, p.~100581, 2020.

\bibitem{tayyab2}
Z.~Yousaf,~M.~Z.~Bhatti and T.~Naseer {\em Eur. Phys. J. Plus},
vol.~135, p.~323, 2020.

\bibitem{PhysRevD.86.127504}
S.~Capozziello, T.~Harko, T.~S. Koivisto, F.~S.~N. Lobo, and G.~J.
Olmo {\em
  Phys. Rev. D}, vol.~86, p.~127504, 2012.

\bibitem{PhysRev.136.B571}
C.~W. Misner and D.~H. Sharp {\em Phys. Rev.}, vol.~136, p.~B571,
1964.

\bibitem{PhysRev.35.875}
R.~C. Tolman {\em Phys. Rev.}, vol.~35, p.~875, 1930.

\bibitem{darmois1927memorial}
G.~Darmois {\em Gauthier-Villars, Paris}, vol.~25, 1927.

\bibitem{herrera1997local}
L.~Herrera and N.~O. Santos {\em Phys. Rep.}, vol.~286, p.~53, 1997.

\bibitem{herrera2009structure}
L.~Herrera, J.~Ospino, A.~Di~Prisco, E.~Fuenmayor, and O.~Troconis
{\em Phys.
  Rev. D}, vol.~79, p.~064025, 2009.

\bibitem{s2} H.~Abedi, A.~M.~Abbassi, S.~Capozziello {\em Ann. Phys-New York}, vol.~405, p.~54, 2019.

\bibitem{s1} S.~Capozziello, C.~ Maurizio, and T.~Maria {\em Ann. Phys-Berlin}, vol.~529, p.~1600376, 2017.

\bibitem{gokhroo1994anisotropic}
M.~K. Gokhroo and A.~L. Mehra {\em Gen. Relativ. Gravit.}, vol.~26,
p.~75, 1994.

\bibitem{herrera2008all}
L.~Herrera, J.~Ospino, and A.~Di~Prisco {\em Phys. Rev. D}, vol.~77,
p.~027502, 2008.

\bibitem{j1} S.~Capozziello,~C.~A.~Mantica, and~L.~G.~Molinari {\em Int. J. Geom. Meth. Mod. Phys.},
vol.~16, p.~1950133, 2019.

\bibitem{j2} S.~Capozziello,~C.~A.~Mantica, and~L.~G.~Molinari {\em Int. J. Geom. Meth. Mod. Phys.},
vol.~16, p.~1950008, 2019.



\end{thebibliography}
\end{document}